\documentclass[12pt]{article}
\usepackage{amsmath,amsfonts,amssymb}
\usepackage{hyperref}
\usepackage{graphicx}
\usepackage{slashed}
\usepackage{color} 
\usepackage{epsfig}
\usepackage{psfrag}
\usepackage{bbm}

\def\TOfootnote#1{ }

\def\be{\begin{equation}}
\def\ee{\end{equation}}
\def\ba{\begin{eqnarray}}
\def\ea{\end{eqnarray}}
\def\bal#1\eal{\begin{align}#1\end{align}}

\def\pic(#1,#2){\begin{figure}[h!]
\begin{center}
  \includegraphics[width=5cm]{#1.png}
  \caption{#2}
 \end{center}
 \end{figure}\\}
 
 \def\mpic(#1,#2){\begin{figure}[h!]
\begin{center}
  \includegraphics[width=7.5cm]{#1.png}
  \caption{#2}
 \end{center}
\end{figure}\\}
 
 \def\bpic(#1,#2){\begin{figure}[h!]
\begin{center}
  \includegraphics[width=10cm]{#1.png}
  \caption{#2}
 \end{center}
\end{figure}\\}
\def\bbpic(#1,#2){\begin{figure}[h!]
\begin{center}
  \includegraphics[width=13cm]{#1.png}
  \caption{#2}
 \end{center}
\end{figure}\\}

\def\spic(#1,#2){\begin{figure}[h!]
\begin{center}
  \includegraphics[width=3cm]{#1.png}
  \caption{#2}
 \end{center}
 \end{figure}\\}

\begin{document}
\bibliographystyle{utphys}

\thispagestyle{empty}

\begin{flushright}
UT-Komaba 18-3
\end{flushright}
\begin{center}
\noindent{{\textbf{\Large 
Interface entropy in four dimensions
\\ \vspace{3mm}
as Calabi's diastasis on the conformal manifold
  }}}\\
\vspace{.5cm}
Kanato Goto and Takuya Okuda
\vspace{0.5 cm}

{\it
 University of Tokyo,
 Komaba,
 Tokyo 153-8902, Japan}\\

\vskip .5cm
\end{center}

\begin{abstract}

We conjecture an equality between (1) the entropy associated with a Janus interface in a 4d $\mathcal{N}=2$ superconformal field theory and (2) Calabi's diastasis, a particular combination of analytically continued K\"ahler potentials, on the conformal manifold (moduli space) of the 4d theory.

\end{abstract}

\vskip .3cm

A Janus interface is a codimension-one object across which coupling constants take different values.
The original construction in 4d~\cite{Bak:2003jk,Clark:2004sb} did not preserve supersymmetry.
This paper concerns half BPS Janus interfaces in 4d $\mathcal{N}=4$~\cite{Gaiotto:2008sd} and $\mathcal{N}=2$~\cite{Drukker:2010jp} superconformal field theories.

It is known~\cite{Bachas:2013nxa} that the interface entropy, or the $g$-factor~\cite{Affleck:1991tk},  of the Janus interface in 2d $\mathcal{N}=(2,2)$ theories is given by a specific combination of analytically continued K\"ahler potentials, namely, Calabi's diastasis~\cite{MR0057000}.
It is natural to ask if a similar statement holds in other dimensions.

Recall that the interface entropy in 2d admits an interpretation as a contribution of the interface to the entanglement entropy for the entangling region containing the interface~\cite{Calabrese:2004eu}.

The authors of~\cite{Estes:2014hka} defined the contribution of an interface (called ``defect'' by them) to the entanglement entropy as follows.
Let us consider the 4d Minkowski space with coordinates $x^\mu$ ($\mu=0,\ldots,3$), together with a conformal interface located at $x^3=0$.
We have CFT$_+$ (CFT$_-$) in the region $x^3>0$ ($x^3<0$).
Take as the entangling region the round two-sphere of radius $R$ centered at the origin, and denote the resulting entanglement entropy by $S$.
Let $S_\pm$ be the entanglement entropy defined by the same geometry, but without an interface, for CFT$_\pm$ respectively.
They define
\begin{equation}
S_\text{interface} = S - \frac{S_+ + S_-}{2} \,.
\end{equation}
In this combination contributions from the bulk region are canceled.
Let $\epsilon$ be a UV cut-off.
By a holographic analysis they found the behavior
\begin{equation}
S_\text{interface}  = D_1 \frac{R}{\epsilon} + D_0 +\mathcal{O}(\epsilon) \,.
\end{equation}
They showed $D_1$ to be scheme dependent, but $D_0$ scheme independent and thus universal.

For the Janus interface, the quantity $D_0$ is a non-holomorphic function $D_0(\tau,\overline{\tau};\tau',\overline{\tau}')$ that depends on two points $\tau$ and $\tau'$ on the conformal manifold.
These points correspond to the values of the couplings on the two sides of the interface.
The function is subject to some constraints.
First, it has to vanish when the two points coincide, ${\it i.e.}$, when the interface is trivial.
Second, it has to be invariant under S-duality transformations.

Another intriguing fact is that the sphere partition functions of 2d $\mathcal{N}=(2,2)$ and 4d $\mathcal{N}=2$ SCFTs both compute the K\"ahler potential $K$ of the respective conformal manifold~\cite{Jockers:2012dk,Gomis:2012wy,Gerchkovitz:2014gta}.
With a suitable normalization of the metric, the relation in the 4d case reads
\begin{equation} \label{ZS4-K}
Z_{S^4}= \left(\frac{r}{ r_0}\right)^{-4a} e^{K/12} \,,
\end{equation}
where $a$ is an anomaly coefficient, $r$ is the radius, and $r_0$ is a renormalization scale.

Any scheme-independent physical quantity associated with a Janus interface should be invariant under duality transformations when there is no globally defined K\"ahler potential~\cite{Donagi:2017vwh}.
Calabi's diastasis (the expression in the bracket of the formula below) possesses this property because it is invariant under K\"ahler transformations.
This makes it a very natural function to enter the universal part of the interface entropy.

These considerations motivate us to make the following conjecture.

\begin{center}
\begin{minipage}{13cm}
{\it
There exists an appropriate notion of interface entropy $D_0$ generalizing the definition above such that for a half BPS Janus interface in a 4d $\mathcal{N}=2$ superconformal field theory $D_0$ is proportional to Calabi's diastasis on the conformal manifold:
\begin{equation} \label{D0-conjecture}
D_0 =c_0 \left[K(\tau,\overline \tau)+ K(\tau',\overline \tau') - K(\tau,\overline \tau') - K(\tau', \overline \tau)\right] \,,
\end{equation}
where $c_0$ is a constant.
}
\end{minipage}
\end{center}

The paper~\cite{Estes:2014hka} applied the holographic formula~\cite{Ryu:2006bv} for entanglement entropy to the supergravity background of~\cite{DHoker:2007zhm} dual to the half BPS Janus interface in 4d $\mathcal{N}=4$ $SU(N)$ super Yang-Mills theory.
Their result for $ \theta^+=\theta^- $, found in  equation (3.72) of their paper,
%
is that
\begin{equation} \label{D0-holographic}
D_0 = - \frac{N^2}{2} \log \left( 1+ \frac{(g_{\rm YM}^+-g_{\rm YM}^-)^2}{2 g_{\rm YM}^+g_{\rm YM}^{ -}} \right)  \,,
\end{equation} 
where $g^\pm_{\rm YM}$ and $\theta^\pm$ are the values of the Yang-Mills coupling and the theta parameter on the two sides of the interface.
Although this formula is written for the special case $\theta^+=\theta^-$, the result for the general case where both $g_{\rm YM}$ and $\theta$ vary across the interface can be obtained by an action of $SL(2,\mathbb{R})$.
This group is a symmetry of type IIB supergravity and transforms the dilaton and the RR scalar that are related to CFT parameters as $C_{(0)}=\theta/2\pi$, $e^{-2\phi} = 4\pi/g_{\rm YM}^2$.
Then the result (\ref{D0-holographic}) of \cite{Estes:2014hka} can be summarized by saying that 
$
c_0= - 1/24
$
with
\begin{equation} \label{K_N=4}
K=  -6 N^2\log i(\overline\tau-\tau)\,,
\qquad
\tau = \frac{\theta}{2\pi} + i \frac{4\pi}{g_{\rm YM}^2}
\end{equation}
 in the large $N$ limit.
The K\"ahler potential~(\ref{K_N=4}) gives the usual metric of constant negative curvature on the upper half plane, which is known to be the Zamolodchikov metric of the theory~\cite{Papadodimas:2009eu}.
 The normalization of $K$ is determined by the relation~(\ref{ZS4-K}) and the coupling dependence $Z_{S^4} \propto g_{\rm YM}^{N^2-1}$~\cite{Pestun:2007rz}.
 This provides a modest check of our conjecture.
 
The appearance of $N^2$ ($\sim$ central charge) in~(\ref{D0-holographic}) is similar to the situation of \cite{DHoker:2014qtw}, where the holographic computation of the interface entropy for certain 2d CFT's yielded the product of the central charge and Calabi's diastasis on the moduli space of the dual supergravity.

More study is desired to check, prove, or generalize the conjecture.
For a further check one may construct the holographic dual of the Janus interface for class S theories by deforming supergravity solutions of~\cite{Gaiotto:2009gz}.
Does a relation similar to~(\ref{D0-conjecture}) hold for other quantities such as those characterizing reflection/transmission?
We leave these matters for the future.

\vspace{-4mm}

\section*{Acknowledgements}

\vspace{-2mm}

The research of KG is supported in part by the JSPS Research Fellowship for Young Scientists.
The research of TO is supported in part by the JSPS Grants-in-Aid for Scientific Research No. 16K05312. 
TO thanks C.~Bachas, E.~D'Hoker, and T.~Nishioka for useful discussion and correspondence.
TO also acknowledges the Galileo Galilei Institute for Theoretical Physics for the hospitality and the INFN for partial support during the completion of this work.

\vspace{-5mm}

\bibliography{refs}

\providecommand{\href}[2]{#2}\begingroup\raggedright\begin{thebibliography}{10}

\bibitem{Bak:2003jk}
D.~Bak, M.~Gutperle, and S.~Hirano, ``{A Dilatonic deformation of AdS(5) and
  its field theory dual},''
  \href{http://dx.doi.org/10.1088/1126-6708/2003/05/072}{{\em JHEP} {\bfseries
  05} (2003) 072},
\href{http://arxiv.org/abs/hep-th/0304129}{{\ttfamily arXiv:hep-th/0304129
  [hep-th]}}.

\bibitem{Clark:2004sb}
A.~B. Clark, D.~Z. Freedman, A.~Karch, and M.~Schnabl, ``{Dual of the Janus
  solution: An interface conformal field theory},''
  \href{http://dx.doi.org/10.1103/PhysRevD.71.066003}{{\em Phys. Rev.}
  {\bfseries D71} (2005) 066003},
\href{http://arxiv.org/abs/hep-th/0407073}{{\ttfamily arXiv:hep-th/0407073
  [hep-th]}}.

\bibitem{Gaiotto:2008sd}
D.~Gaiotto and E.~Witten, ``{Janus Configurations, Chern-Simons Couplings, And
  The theta-Angle in N=4 Super Yang-Mills Theory},''
  \href{http://dx.doi.org/10.1007/JHEP06(2010)097}{{\em JHEP} {\bfseries 06}
  (2010) 097},
\href{http://arxiv.org/abs/0804.2907}{{\ttfamily arXiv:0804.2907 [hep-th]}}.

\bibitem{Drukker:2010jp}
N.~Drukker, D.~Gaiotto, and J.~Gomis, ``{The Virtue of Defects in 4D Gauge
  Theories and 2D CFTs},''
  \href{http://dx.doi.org/10.1007/JHEP06(2011)025}{{\em JHEP} {\bfseries 06}
  (2011) 025},
\href{http://arxiv.org/abs/1003.1112}{{\ttfamily arXiv:1003.1112 [hep-th]}}.

\bibitem{Bachas:2013nxa}
C.~P. Bachas, I.~Brunner, M.~R. Douglas, and L.~Rastelli, ``{Calabi's diastasis
  as interface entropy},''
  \href{http://dx.doi.org/10.1103/PhysRevD.90.045004}{{\em Phys. Rev.}
  {\bfseries D90} no.~4, (2014) 045004},
\href{http://arxiv.org/abs/1311.2202}{{\ttfamily arXiv:1311.2202 [hep-th]}}.

\bibitem{Affleck:1991tk}
I.~Affleck and A.~W.~W. Ludwig, ``{Universal noninteger 'ground state
  degeneracy' in critical quantum systems},''
\href{http://dx.doi.org/10.1103/PhysRevLett.67.161}{{\em Phys. Rev. Lett.}
  {\bfseries 67} (1991) 161--164}.

\bibitem{MR0057000}
E.~Calabi, ``Isometric imbedding of complex manifolds,''
  \href{http://dx.doi.org/10.2307/1969817}{{\em Ann. of Math. (2)} {\bfseries
  58} (1953) 1--23}. \url{https://doi.org/10.2307/1969817}.

\bibitem{Calabrese:2004eu}
P.~Calabrese and J.~L. Cardy, ``{Entanglement entropy and quantum field
  theory},'' \href{http://dx.doi.org/10.1088/1742-5468/2004/06/P06002}{{\em J.
  Stat. Mech.} {\bfseries 0406} (2004) P06002},
\href{http://arxiv.org/abs/hep-th/0405152}{{\ttfamily arXiv:hep-th/0405152
  [hep-th]}}.

\bibitem{Estes:2014hka}
J.~Estes, K.~Jensen, A.~O'Bannon, E.~Tsatis, and T.~Wrase, ``{On Holographic
  Defect Entropy},'' \href{http://dx.doi.org/10.1007/JHEP05(2014)084}{{\em
  JHEP} {\bfseries 05} (2014) 084},
\href{http://arxiv.org/abs/1403.6475}{{\ttfamily arXiv:1403.6475 [hep-th]}}.

\bibitem{Jockers:2012dk}
H.~Jockers, V.~Kumar, J.~M. Lapan, D.~R. Morrison, and M.~Romo, ``{Two-Sphere
  Partition Functions and Gromov-Witten Invariants},''
  \href{http://dx.doi.org/10.1007/s00220-013-1874-z}{{\em Commun. Math. Phys.}
  {\bfseries 325} (2014) 1139--1170},
\href{http://arxiv.org/abs/1208.6244}{{\ttfamily arXiv:1208.6244 [hep-th]}}.

\bibitem{Gomis:2012wy}
J.~Gomis and S.~Lee, ``{Exact Kahler Potential from Gauge Theory and Mirror
  Symmetry},'' \href{http://dx.doi.org/10.1007/JHEP04(2013)019}{{\em JHEP}
  {\bfseries 04} (2013) 019},
\href{http://arxiv.org/abs/1210.6022}{{\ttfamily arXiv:1210.6022 [hep-th]}}.

\bibitem{Gerchkovitz:2014gta}
E.~Gerchkovitz, J.~Gomis, and Z.~Komargodski, ``{Sphere Partition Functions and
  the Zamolodchikov Metric},''
  \href{http://dx.doi.org/10.1007/JHEP11(2014)001}{{\em JHEP} {\bfseries 11}
  (2014) 001},
\href{http://arxiv.org/abs/1405.7271}{{\ttfamily arXiv:1405.7271 [hep-th]}}.

\bibitem{Donagi:2017vwh}
R.~Donagi and D.~R. Morrison, ``{Conformal field theories and compact curves in
  moduli spaces},''
\href{http://arxiv.org/abs/1709.05355}{{\ttfamily arXiv:1709.05355 [hep-th]}}.

\bibitem{Ryu:2006bv}
S.~Ryu and T.~Takayanagi, ``{Holographic derivation of entanglement entropy
  from AdS/CFT},'' \href{http://dx.doi.org/10.1103/PhysRevLett.96.181602}{{\em
  Phys. Rev. Lett.} {\bfseries 96} (2006) 181602},
\href{http://arxiv.org/abs/hep-th/0603001}{{\ttfamily arXiv:hep-th/0603001
  [hep-th]}}.

\bibitem{DHoker:2007zhm}
E.~D'Hoker, J.~Estes, and M.~Gutperle, ``{Exact half-BPS Type IIB interface
  solutions. I. Local solution and supersymmetric Janus},''
  \href{http://dx.doi.org/10.1088/1126-6708/2007/06/021}{{\em JHEP} {\bfseries
  06} (2007) 021},
\href{http://arxiv.org/abs/0705.0022}{{\ttfamily arXiv:0705.0022 [hep-th]}}.

\bibitem{Papadodimas:2009eu}
K.~Papadodimas, ``{Topological Anti-Topological Fusion in Four-Dimensional
  Superconformal Field Theories},''
  \href{http://dx.doi.org/10.1007/JHEP08(2010)118}{{\em JHEP} {\bfseries 08}
  (2010) 118},
\href{http://arxiv.org/abs/0910.4963}{{\ttfamily arXiv:0910.4963 [hep-th]}}.

\bibitem{Pestun:2007rz}
V.~Pestun, ``{Localization of gauge theory on a four-sphere and supersymmetric
  Wilson loops},'' \href{http://dx.doi.org/10.1007/s00220-012-1485-0}{{\em
  Commun. Math. Phys.} {\bfseries 313} (2012) 71--129},
\href{http://arxiv.org/abs/0712.2824}{{\ttfamily arXiv:0712.2824 [hep-th]}}.

\bibitem{DHoker:2014qtw}
E.~D'Hoker and M.~Gutperle, ``{Holographic entropy and Calabi`s diastasis},''
  \href{http://dx.doi.org/10.1007/JHEP10(2014)093}{{\em JHEP} {\bfseries 10}
  (2014) 093},
\href{http://arxiv.org/abs/1406.5124}{{\ttfamily arXiv:1406.5124 [hep-th]}}.

\bibitem{Gaiotto:2009gz}
D.~Gaiotto and J.~Maldacena, ``{The Gravity duals of N=2 superconformal field
  theories},'' \href{http://dx.doi.org/10.1007/JHEP10(2012)189}{{\em JHEP}
  {\bfseries 10} (2012) 189},
\href{http://arxiv.org/abs/0904.4466}{{\ttfamily arXiv:0904.4466 [hep-th]}}.

\end{thebibliography}\endgroup
 \end{document}